# Variable range hopping conduction in semiconductor nanocrystal solids

Dong Yu, Congjun Wang, Brian L. Wehrenberg, Philippe Guyot-Sionnest
*James Franck Institute, University of Chicago, Chicago Illinois 60637*

**The temperature and electrical field dependent conductivity of n-type CdSe nanocrystal thin films is investigated. In the low electrical field regime, the conductivity follows $\sigma \sim \exp(-(T^*/T)^{1/2})$ in the temperature range 10K<T<120K. At high electrical field, the conductivity is strongly field dependent. At 4K, the conductance increases by eight orders of magnitude over one decade of bias. At very high field, conductivity is temperature-independent with $\sigma \sim \exp(-(E^*/E)^{1/2})$. The complete behavior is very well described by variable range hopping with Coulomb gap.**

Electron transport in amorphous semiconductors has been investigated for decades [1,2]. In recent years, "artificial solids" formed by close packed arrays of monodispersed metallic nanostructures have provided novel model systems for the study of electron transport [3, 4, 5]. Semiconductor nanocrystals [6] can now be synthesized for a wide range of materials with nanometer sizes leading to strong quantum confinement. One of the best characterized systems is CdSe [7]. In several studies, neutral CdSe nanocrystal solids have been found to be photoconductive [8,9] but highly insulating [10]. Charging of the close-packed assemblies of the CdSe quantum dots (QDs) leads to vastly increased conductivity [11]. In this paper, the electronic transport mechanism of CdSe n-type semiconductor nanocrystal solids is studied through temperature and field dependent dc-conduction experiments.

Because of the lack of global structural order, the n-type semiconductor nanocrystal solids can be considered as an Anderson insulator. Electrons are localized close to the Fermi level but an electron could hop from one localized site to another when receiving energy from a phonon or directly from external electrical field. Mott first pointed out that at low temperature the most frequent hopping would not be to the nearest neighbor. Mott's Variable Range Hopping model (M-VRH) gives the relation between conductance $G$ and temperature $T$ as [1],

$$G \propto \exp(-B/T^\nu) \tag{1}$$

with $\nu = 1/4$ for a 3-D system, $\nu = 1/3$ for 2-D. The essence of VRH can be summarized as below. A hopping electron will always try to find the lowest activation energy $\Delta E$ and the shortest hopping distance. However, usually the two conditions cannot be satisfied at the same time. Thus there will be an optimum hopping distance $r$, which maximizes the hopping probability. In zero bias condition, the probability is,

$$P \sim \exp(-2r/a - \Delta E/k_B T) \tag{2}$$

where $a$ is the localization length, $k_B$ is the Boltzmann constant and $\Delta E$ is the activation energy.

Taking $\Delta E \sim 1/g_0 r^3$ (where $g_0$ is the density of states) and assuming a constant $g_0$ at the Fermi level, Mott derived a $\nu = 1/4$ law [1]. Considering the Coulomb interaction between the charged sites, Efros and Shklovskii (ES-VRH) showed that $g_0$ vanishes quadratically at the Fermi level and the conductance follows a $\nu = 1/2$ law [12]. In this model, the activation energy $\Delta E$ is proportional to $1/r$ as in the Coulomb potential and it has a positive sign. In this Letter, we simply call $\Delta E$ "Coulomb barrier". However, this term is not the conventional Coulomb potential because it comes from a many body problem. It differs numerically from a Coulomb potential by a factor which was determined by computer simulation [2].

Above a certain critical temperature $T_C$, the Coulomb interaction can be neglected and the conductivity obeys Mott's law ($\nu = 1/4$) while below $T_C$ the Efros-Shklovskii's $\nu = 1/2$ law holds [12]. $T_C$ is given by

$$T_C = \frac{e^4 a g_0}{k_B (4\pi\varepsilon\varepsilon_0)^2} \tag{3}$$

$T_C$ is very small for usual amorphous materials. According to Knotek *et al.* [13], for amorphous Ge, $g_0 = 1.5 \times 10^{18}$ eV$^{-1}$cm$^{-3}$, $a = 1$ nm, which gives $T_C = 0.15$ K. In highly monodispersed semiconductor nanocrystals quantum confinement leads to discrete energy states with narrow homogeneous linewidth. If the overall linewidth is 100 meV and the radius of the QDs is ~3 nm, the density of states $g_0 \sim 9 \times 10^{19}$ eV$^{-1}$cm$^{-3}$ is much larger than that for amorphous Ge. In QD arrays, the localization length is estimated by the radius of the dots which is ~ 3 nm. Using an effective dielectric constant $\varepsilon \sim 4$ for the close-packed CdSe QD films [14], $T_C$ is estimated to be about 400 K in our samples, which means that the Coulomb gap plays an important role and the $v = 1/2$ law is expected.

The diameter of the nanocrystals is ~5.4 nm with a size variance of < 5% and the inter-dot distance is ~0.7 nm. The nanocrystal films are made by drop-casting pyridine capped CdSe QDs solution on a Pt interdigitated working electrode with a 5-micron separation [11] corresponding to about 1000 diameters. The thickness of the film ranges from 10 to 50 layers of QDs and is determined by the optical density of the film. The film is then cross-linked by 1,4-phenylenediamine and baked under $N_2$ at 70 ºC for 2 hours. This treatment helps achieve fast and robust charging of QDs [14]. The sample is further dried in vacuum overnight then immersed in anhydrous electrolyte solution (e.g. 0.1 M tetrabutylammonium tetrafluoroborate in *N,N*-dimethylformamide), which is sealed inside an electrochemical cell under nitrogen.

The CdSe nanocrystals are reduced by setting the working electrodes at a negative potential relative to a silver pseudo-reference electrode [11]. Electrons are injected into the lowest quantum state, $1S_e$, as determined by optical measurements [11, 14, 15]. The conductance is measured by applying a fixed small bias between the two working electrodes. The sample is held at the negative potential while the electrochemical cell is cooled down in a liquid helium storage dewar. The temperature is read from a silicon diode embedded in the Teflon body of the electrochemical cell. When the electrolyte solution freezes (~ 200K), the bipotentiostat is disconnected from the sample and the QD solids remain charged. The film conductance is stable for days when the sample is kept below ~ 160 K. At the same temperature, the *IV* curves are always the same no mater how long the sample stays at low temperature and how many times the sample is warmed up and cooled down. The frozen electrolyte solution doesn't contribute to the conduction. As a blank experiment, we measured the conductance of frozen electrolyte solution with the same interdigitated electrodes, which is undetectable (<1 pS) at all temperatures below 160 K.

*IV* curves are measured after the cell reaches thermal equilibrium. The *IV* curves are highly symmetric at all temperatures (Fig. 1) and show no hysteresis at the ramp rates used (~100 mV/s). The conductance drops as the temperature decreases and the *IV* curves are quite linear above 30 K. Below about 20 K, voltage thresholds appear above which the *IV* curves are extremely nonlinear (Fig. 3B).

To study the temperature dependent conductance at low bias, the film is cooled to 4 K then warmed up slowly (temperature increases at a rate of about 1 K/min) to maintain thermal equilibrium. The current is measured at a small bias of 1 V. The current is too small to detect (<1 pA) below 11 K and the electrolyte solution starts to melt above 160 K. Between 11 K and 120 K the conductance $G$ (defined as $G = I/V$) follows very well the $v = 1/2$ law (Fig. 2)

$$G = A\exp(-\sqrt{\frac{T^*}{T}}) \tag{4}$$

The constant $T^*$ is ~5.2 × 10$^3$ K and $A$ is ~9.8 × 10$^{-3}$ S from the data fitting. The prefactor $A$ includes tunneling through the high barrier created by the organic capping layers, and accounts for the attempt frequency of electrons trying to escape the nanocrystal.

$T^*$ is in fact predicted by the ES-VRH model [2],

$$T^* \approx \frac{2.8e^2}{4\pi\varepsilon\varepsilon_0 a k_B} \tag{5}$$

Using the same values for the parameters as in the calculation of $T_C$, we obtained an expected $T^* = 3.9 \times 10^3$ K, which is in excellent qualitative agreement with the experimental value. ES-VRH over a smaller temperature range and with an even higher $T^*$ has been recently reported for nanocrystalline $SiO_2/CdTe/SiO_2$ semiconductor films [16].

Above 120 K, the temperature dependence of the conductance starts to follow Arrhenius behavior (Fig. 2 inset). At these higher temperatures, Arrhenius behavior was previously reported in this system [11] as well as in ZnO nanocrystals assemblies [17]. In the VRH model, this happens when $T$ is larger than $T_A$ where nearest neighbor hopping is most favored. $T_A = \left(\frac{a}{4d}\right)^2 T^*$ and $d$ is the nearest neighbor distance. Using $T_A = 120$ K, one determines that the localization length is $a \sim 3.6$ nm which is consistent with the earlier estimate. Achieving longer localization length will require reducing the barrier heights between the nanocrystals. Above $T_A$ the conductance becomes thermally activated as $G = A\exp(-\frac{\Delta E_A}{k_B T})$ where the Arrhenius activation energy is $\Delta E_A = \frac{a}{8d}k_B T^*$ = 33 meV. $\Delta E_A$ is about 30 meV as determined by data fitting indicating that the model is fully internally consistent. This is significantly smaller than the expected charging energy [18]. Indeed, using Eq. (5), $\Delta E_A \approx 0.35 E_C$, where $E_C = \frac{e^2}{4\pi\varepsilon\varepsilon_0 d}$ is a simple estimate of the charging energy.

The average number of electrons per dot was lowered by briefly melting the electrolyte at about 200 K. The film conductance dropped ten times after this partial discharge but the temperature dependence still followed ES-VRH (Fig. 2 lower curve). The parameter $T^*$ is a little larger ($\sim 6.2 \times 10^3$ K) in this case. The difference of $T^*$ is probably due to the smaller localization length and reduced screening at lower doping level.

As discussed above, ES-VRH fits extremely well the data at low bias. However at high bias the temperature dependence is strongly reduced, and one enters a regime where electron transport is field driven. The conductance increases by 8 orders of magnitude over one decade of bias at 4.3 K (Fig. 3A). This extremely large dynamic range suggests the following modification of the hopping model. We consider that the down field energy is now reduced by *eEr* due to the electrical field *E*. Electron hopping will be much facilitated when the voltage drop is of the order of the Coulomb barrier. We now write the hopping conductance to a nanocrystal downstream at a distance *r* as,

$$G = A\exp(-2\frac{r}{a} - \frac{T^*}{T}\frac{a}{8r} + \frac{eEr}{k_BT}) \tag{6}$$

When *E* is larger than a critical field $E_C = \frac{2k_BT}{ea}$ (e.g $E_C = 2\times10^5$ V/m at 4 K), the conduction is field dominated. In this regime, the external field helps electron overcome the Coulomb barrier, i.e. the last two terms in equation (6) cancel each other. This gives [19, 20] a temperature independent conductance *G*,

$$G = A\exp(-\sqrt{\frac{E^*}{E}}) \text{ where } E^* = \frac{k_BT^*}{2ea} \tag{7}$$

Figs. 3A shows the experimental field-dependence of the conductance at different temperatures. At very low bias, conduction is ohmic and the temperature dependence of the conductance obeys ES-VRH. At higher bias, the conductance is strongly nonlinear and approaches Eq. (7). Using the same parameters as earlier, we calculated $E^* \sim 7.5 \times 10^7$ V/m. This is again in qualitative agreement with the experimental value of $3.4 \times 10^8$ V/m determined by fitting the data at the lowest temperature.

This nonlinearity is not likely due to self-heating effect. We estimated the temperature difference across the thin glass substrate with thickness *l* (~2mm) by,

$$\Delta T = \frac{Pl}{\kappa S} \tag{8}$$

where *P* is the thermal power (<1mW), $\kappa$ is the thermal conductivity of the glass substrate (~2.5 mW/cm·K) and *S* is the electrodes area (~ 10 mm$^2$), we calculated the maximum temperature difference $\Delta T \sim 1$ K. Within the experimental temperature range (*T* > 4 K), this difference can be neglected. In the measurements, the *IV* curves are independent of bias scan rate and the rising and falling curves overlap very well. These observations provide supporting evidence that the self-heating effect is indeed insignificant.

To obtain a model which holds at all temperatures and field strengths, we modified equation (6) by normalizing the Boltzmann term by the partition function as below,

$$G = A \frac{\exp(-2\frac{r}{a} - \frac{T^*}{T}\frac{a}{8r} + \frac{eEr}{k_B T})}{1 + \exp(-\frac{T^*}{T}\frac{a}{8r} + \frac{eEr}{k_B T})} \tag{9}$$

with the same parameters as before.

At small $E$ and $T$, equation (9) simplifies to (2) which leads to the $v = 1/2$ law and the experimental result at low bias (Fig. 2). When $E$ becomes large, electrons will most probably hop to the nearest neighbors and $r = d$ where d is the nearest neighbor distance. This tends to a temperature independent conductance limit $A\exp(-2d/a)$.

The model is then tested by simulations, in which we scanned $r$ to maximize the conductivity in Eq. (9). Figs. 3C-3D demonstrate the resulting conductivity as a function of $T$, $E$ and Figs. 3E-3F illustrate the hopping distance $r$ scaled by the nearest neighbor distance $d$ where the hopping distance first increases then decreases as the field becomes larger. This can be understood from Eq. (6). A small electrical field helps electrons to hop further by reducing the effective activation energy while a large field provides enough energy to overcome the Coulomb barrier such that the electrons only jump to the nearest neighbors. The model reproduces very well the qualitative trend of the experimental data (Figs. 3A-3D) over many orders of magnitude by using the values of $A$, $a$, $T^*$ experimentally determined as above. The details are not exactly the same. The critical field strength calculated is about 4 times weaker than the experimental value. Another difference is that the cross-over between the ohmic regime at low field and the temperature independent regime at high field is somewhat softer in the experimental data than in the simulation (Fig. 3A and 3C). However, given the simplicity of the model, the description is already surprisingly good.

CdSe nanocrystals with trioctylphosphine oxide (TOPO) capping molecules show similar conduction behavior although the conductivity is much smaller possibly because the length of TOPO ligand is longer than that of pyridine. The inter-dot distance increases from ~0.7 nm for pyridine capped QDs to ~1.1 nm in the case of TOPO capped QDs. We also doped the nanocrystal thin films n-type by potassium evaporation in an ultra-high vacuum system [11]. The low temperature *IV* curves showed similar temperature and field dependent conductivity.

In summary, n-type semiconductor monodispersed QD solids exhibit very clearly variable range hopping where the energy barrier is dominated by the Coulomb interaction rather than by energy inhomogeneity. Over a wide temperature range (4 K – 120 K) and fields up to $10^7$ V/m, the variable range hopping model, with only three parameters, provides an excellent qualitative description of the conduction behavior over more than eight orders of magnitude in current. Having developed an understanding for the CdSe system, parameters can now be identified and modified to reduce the temperature dependence of the conductivity and approach a more metallic behavior. Screening is one such experimentally adjustable parameter. Semiconductor nanocrystals made of materials with much larger dielectric constants should exhibit much weaker temperature dependence and possibly higher conductivity.

We acknowledge helpful discussions with H. M. Jaeger, I. Gruzberg and T. F. Rosenbaum. D.Y. was supported by the University of Chicago-Argonne National Laboratory Consortium for Nanoscience Research. C.W. and B.L.W. were supported by the U.S. National Science Foundation (NSF) under grant DMR-0108101. Use of the MRSEC Shared Facilities was supported by NSF under grant DMR-0213745.

[1] N. F. Mott, J. Non-Cryst. Solids **1**, 1 (1968); N. F. Mott, *Conduction in Non-Crystalline materials 2$^{nd}$ Edition* (Clarendon Press, Oxford 1993).
[2] B.I. Shklovskii and A. L. Efros, *Electronic Properties of Doped Semiconductors* (Springer-Verlag, Berlin, 1984).
[3] R. C. Doty, H. Yu, C. K. Shih, and B. A. Korgel, J. Phys. Chem. B **105**, 8291 (2001).

[4] K. C. Beverly, J. F. Sampaio, and J. R. Heath, J. Phys. Chem. B **106**, 2131 (2002).
[5] R. Parthasarathy, X. M. Lin, and H. M. Jaeger, Phys. Rev. Lett. **87**, 186807 (2001).
[6] A. L. Efros and A. L. Efros, Sov. Phys. Semicond. **16**, 772 (1982).
[7] C. B. Murray, C. R. Kagan, and M. G. Bawendi, Annu. Rev. Mater. Sci. **30**, 545 (2000).
[8] C. A. Leatherdale, *et al.*, Phys. Rev. B **62**, 2669 (2000).
[9] D. S. Ginger and N. C. Greenham, J. Appl. Phys. **87**, 1361 (2000).
[10] N. Y. Morgan, *et al.*, Phys. Rev. B **66**, 075339 (2002).
[11] D. Yu, C. Wang, and P. Guyot-Sionnest, Science **300**, 1277 (2003).
[12] A. L. Efros and B. I Shklovskii, J. Phys. C **8**, L49 (1975).
[13] M. L. Knotek *et al.*, Phys. Rev. Lett. **30**, 853 (1973).
[14] P. Guyot-Sionnest and C. Wang, J. Phys. Chem. B **107**, 7355 (2003).
[15] C. Wang, M. Shim, and P. Guyot-Sionnest, Science **290**, 2391 (2001).
[16] S.K. Bera, S. Chauduri, and A. K. Pal, Thin Sol. Films 415, 68 (2002)
[17] A. L. Roest, J. J. Kelly, and D. Vanmaekelbergh, Appl. Phys. Lett. **83**, 5530 (2003).
[18] A. Franceschetti, A. Williamson, and A. Zunger, J. Phys. Chem. B, **104**, 3398 (2000).
[19] N. Apsley and H. P. Hughes, Phil. Mag., **31**, 1327 (1975).
[20] A.V. Dvurechenskii, V. A. Dravin, and A. I. Yakimov, JETP Lett. **48,** 155 (1988).

Figure Caption:

FIG. 1. Current-voltage (*IV*) curves for a typical n-type CdSe nanocrystal thin film. Above 30 K, *IV* curves are close to linear. Conductivity decreases as temperature decreases.

FIG. 2. Temperature dependence of the low-field conductance (bias = 1 V) in the range 11 K < $T$ < 160 K. The two curves are the same sample at different charging levels. From fitting the data between 11K and 120K, $T^*$ = 5.2 (circle), 6.2 (cross) $\times 10^3$ K. The inset plot is ln$G$ vs. 1/$T$ (11 K < $T$ < 160 K), where the conduction clearly deviates from Arrhenius behavior at low temperature.

FIG. 3. Field dependence of conductivity. Solid lines are experimental data and dashes lines are from simulation. (A-B) Experimental results with field strength range from $10^5$ V/m to $10^7$ V/m. By fitting the linear part of the curve at 4.3 K in (A), $E^*$= 3.4 × $10^8$ V/m. (C-F) Simulation based on normalized variable range hopping model. $r/d$ is the hopping distance scaled by the nearest neighbor distance $d$. For the simulation, the values of $A$ = $9.8 \times 10^{-3}$ S, $T^*$ = 5200 K and $a$ = 3.6 nm from low field experiments are used.

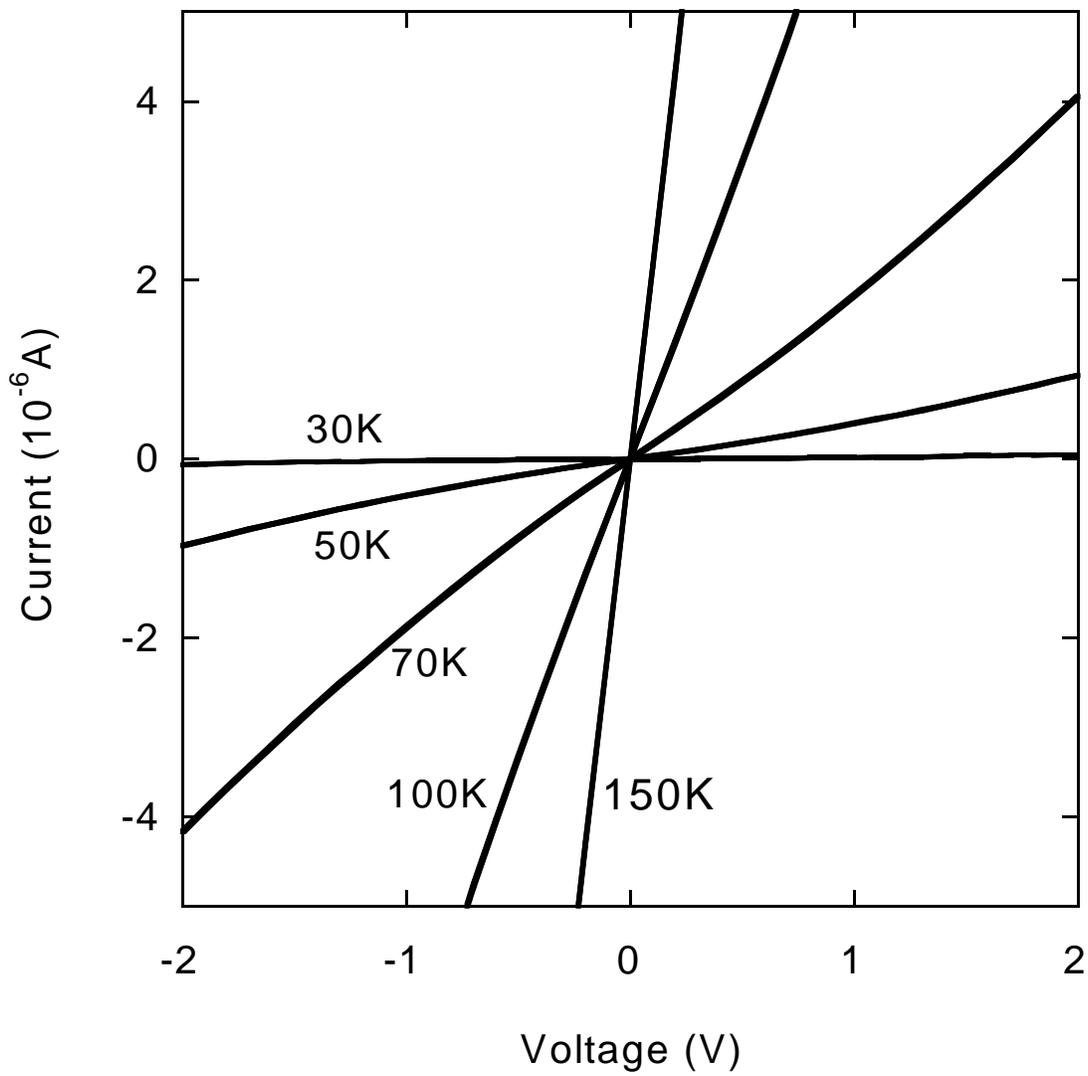

FIG. 1.

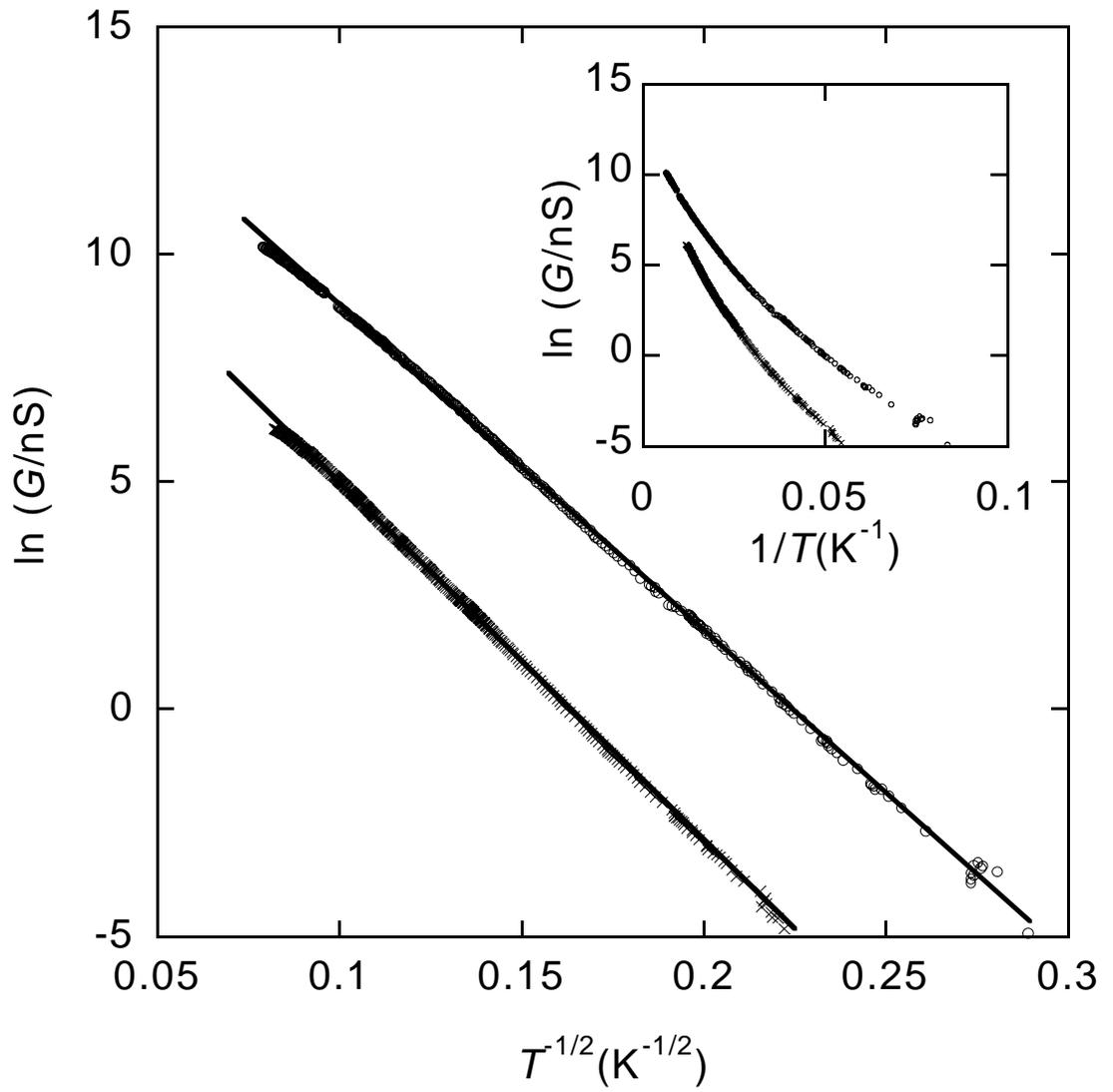

FIG. 2.

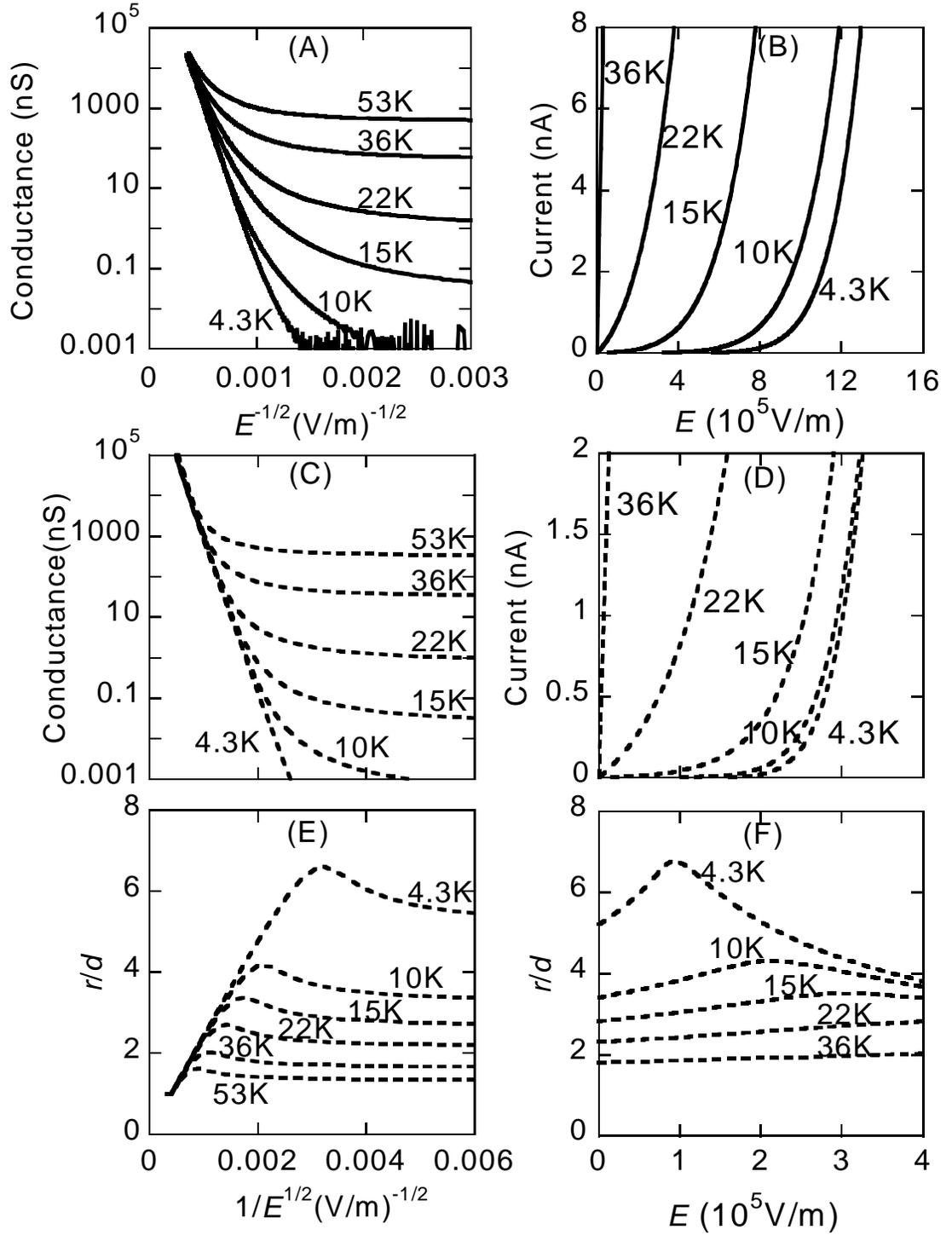

FIG. 3.